\documentclass[14pt]{article}
\usepackage{amsmath,amssymb,graphicx,mathrsfs,hyperref,slashed}  

\begin{document}

\title{\vspace{-1.8in}
\vspace{0.3cm} {Quelling the concerns \\ of \\ EPR and Bell \\}}

\author{\large K.L.H. Bryan,$\;\;\;\;\;$  A.J.M. Medved\\
 \hspace{-1.5in} \vbox{
 \begin{flushleft}
$^{\textrm{\normalsize   Department of Physics \& Electronics, Rhodes University, Grahamstown 6140, South Africa }}$
 \\ \small \hspace{1.7in}
     g08b1231@gmail.com,$\;\;\;\;\;$j.medved@ru.ac.za
\end{flushleft}
}}
\date{}
\maketitle

\begin{abstract}
 
\noindent
We begin with a review of the famous thought experiment that was proposed by Einstein, Podolsky and Rosen (EPR) and mathematically formulated  by Bell; the
outcomes of which challenge the completeness of quantum mechanics and the locality of Nature. We then suggest a reinterpretation of the EPR experiment that 
utilizes  observer complementarity; a concept from quantum gravity which allows spatially separated observers to have their own, independent reference frames. 
The resulting picture provides a self-consistent resolution of the situation that does not jeopardize causality nor unitarity, nor does it resort to ``spooky'' 
(non-local) interactions. Our conclusion is that EPR and Bell rely on an overly strong definition of locality that is in conflict with fundamental physics.

\end{abstract}

\newpage

\tableofcontents

\section{Introduction}

\subsection{Prologue}

Quantum mechanics is a field of study that is infested with
counter-intuitive concepts, and many of our classical
preconceptions are brought into question when we deal with situations in the 
quantum realm.

At the forefront of quantum paradoxes is the thought experiment that was
 put forward in the famous and influential paper by Einstein, Podolsky and Rosen (EPR).
The EPR paper sparked a debate that is still in progress and  requires
a resolution if quantum mechanics is to be validated as a  theory which gives a complete description of reality
in its domain of applicability.

This paper examines the EPR argument as well as
Bell's formulation thereof  and then,  using
a concept brought in from quantum gravity, suggests a different
tact for addressing  the unease of the situation.
In brief, by reworking the EPR and Bell scenario within the framework of 
observer complementarity, any ``spooky
      action at a distance'' is vanquished and the concerns of EPR 
and Bell are  resolved.

\subsection{Where it all started}

The EPR thought experiment presented a challenge for quantum mechanics. It highlighted a 
fundamental paradox within the theory which suggested that, while the theory worked as a 
calculation device, it may not represent reality.
      
      In the EPR paper \cite{EPR}, the argument begins with the consideration of
      a standard quantum-mechanical concept; namely, non-commuting operators.
      Let us consider an experimenter, called Alice, who has access to a particle in a state given by $|\Psi\rangle$.
The mathematical description of Alice physically measuring a property of
      the particle, say momentum, is given by acting on the initial state with
      the relevant operator. The value that the operator retrieves is associated
      with the physical value that would be measured by Alice. EPR does not present
      a comprehensive definition of reality but, rather, remains satisfied with
      a ``criterion of reality'' by which they identify a physical reality with 
      the  corresponding physical quantity. By this criterion, 
      the value produced by an operator acting on a state 
      represents  a real physical quantity; an element of reality. 
      
      However, quantum operators are not always so compliant and the issue of  non-commuting operators arises. The momentum and position
of a particle  form just such  a pair of  incompatible  operators.
   The EPR argument then states that, if Alice had measured the position of 
      the particle rather than its momentum,  the position has physical reality
      but the momentum does not. The original state is changed by the measurement interaction
      and the momentum value of the original state is lost as a consequence of measuring the position.
      And so there are pairs of operators for which the corresponding physical 
      values may be determined but having a definite value for one implies that
      no definite value exists for the other.~\footnote{This statement is 
essentially an
      iteration  of the Heisenberg uncertainty principle.}
      By this reasoning, the EPR argument concludes that we must either accept that 
      these physical quantities relating to incompatible observables 
      do exist but quantum mechanics does not fully
      describe them {\em or\/} that the properties associated with two 
      non-commuting operators cannot have physical reality at the same time.
      Only one can be considered an element of reality.
       
      The next step of the EPR argument involves combining the EPR reality
      criterion with the assumption that quantum mechanics does offer a complete
      description. Their logic is perhaps more easily explained in terms of the spins of two
      particles, rather than the position and momentum of a single particle as originally used in the EPR paper.
      Bohm introduced this newer formulation of the EPR setup and showed that the argument's
      structure remained the same, regardless of which pair of non-commuting observables
      are being considered \cite{Bohm}. Bohm's scenario uses a pair of particles in a singlet state
      and the role of non-commuting operators is now adopted by the different
components of either particle's spin vector.
 This state can be described, in standard  Dirac notation,~\footnote{The conventions of this paper are aligned with \cite{Sakurai}.} as
      \begin{align}
\label{singlet}
       |\Psi\rangle=\frac{1}{\sqrt{2}}\Big[|a+\rangle|b-\rangle \;-\; |a-\rangle|b+\rangle\Big]\;,
      \end{align}
      with the two particles, $\alpha$ and $\beta$, having anti-correlated spins of $a\pm$ and $b\mp$, respectively. Here,  $+$ or $-$ refers to spin up or down 
with respect
to some chosen reference axis (typically but not exclusively, the $z$-axis).
      
   We now introduce two spatially separated observers, Alice and Bob.
      They are each sent one particle, $\alpha$ and $\beta$ respectively, from an initial starting point.
      The particles, prior to their being sent off, are prepared 
 in a  singlet state as described above. 
All the information about the particles is provided by
      the relevant theory; presumably,  orthodox quantum mechanics. 
(Later on, $\lambda$ is used to denote, schematically,  the encapsulation of 
all this information.)
     When the particles arrive at Alice and Bob,
      each observer has a choice as to which direction they choose to measure.
In general, they choose directions that differ from the initial reference
axis and differ from one another.

Let us now focus on Alice's measurement of $\alpha$. Alice can determine $\alpha$'s spin
      in one direction only, say $\hat{n}_{a}$. However, the perfect anti-correlation of the 
      particles means that, by measuring the spin in a given direction, 
      Alice will be able to predict $\beta$'s spin in that same direction. By the reality 
      criterion of EPR, this means that $\beta$ has a physical property relating to Alice's measurement value.
      This physical property, as EPR points out, cannot have ``sprung'' into existence
      at Bob's location as a result of Alice's measurement as this violates 
the condition
      of locality \cite{EPR}. A non-local action by Alice cannot
  instantaneously effect Bob.
      It should be noted, though, that this is not a violation of 
      causality  but of the definition of locality as assumed by EPR.
      Although Bob's outcome can depend on the choice made by Alice
(assuming that he, by chance, chooses $\;\hat{n}_b=\hat{n}_a$), she cannot communicate with
      or signal  Bob  using {\em only} the anti-correlation of the particles 
({\em e.g.}, \cite{Ghirardi}). In order to signal Bob, Alice would be required to 
      pick up a phone, write an email or physically move to Bob's location to relay any
      information regarding the chosen direction. 
      The puzzle does not lie, then, in the realm of faster-than-light
      signaling but in the world of ``spooky'' dependence of the state on non-local variables \cite{Mermin3}.
     
	   EPR's contention, therefore, is that the physical quantity 
corresponding to the spin 
	    of  $\beta$  in the  direction $\hat{n}_{a}$  must have been 
	    determined {\em before} the spatial separation of the two spins.
Then,  remembering that Alice
	    could have chosen any direction to measure in, EPR are lead
to the conclusion that
	    {\em all} the physical properties of $\beta$'s spin must have been 
similarly predetermined.
	    This, by their reality criterion, implies a physical reality for 
the spin of $\beta$ in {\em each and  every} direction. And so we must face the contradiction of having a set of  non-commuting operators
	    (the different components of spin) corresponding to simultaneously real properties.
	    
	    As stated above, the EPR argument comes down to  
the following  choice: The inability  of   quantum mechanics
to completely describe physical reality versus  
  the properties of  non-commuting 
	    observables not  simultaneously existing.
But, since the above argument  has shown that assuming
	   the completeness of  quantum mechanics leads to the simultaneous existence of   non-commuting spin directions,  the conclusion
apparently must be  that the quantum mechanics 
provides  an  incomplete description of all  elements of reality.
      Although the EPR paper is very clear about its conclusion, there are those who disagree
      with this outcome. At the forefront of the dissension  is Bohr's interpretation of the EPR experiment \cite{Bohr}.
	   One main concern for Bohr was EPR's assumption that hypothetical experiments
	    could be used in conjunction with experiments that were actually done; for example, considering
	    the value of momentum after measuring the position. Bohr's argument led him to conclude
	    that the inability to describe certain situations is a part of reality. 
	    
	   This assumption of EPR --- that   physical meaning can
be ascribed  to hypothetical measurements  ---  
	   is a large part of the disagreement between Bell's formulation
of the EPR experiment  and the argument that will be put forth  
in  the current paper.
	   The reliance of EPR and Bell's theorem on counter-factual statements about such  hypothetical events make 
	   their outcomes incompatible with {\em observer complementarity}; an important consequence of
	   quantum gravity that may   (and, as later argued, does) 
permeate into more
``conventional''  physics.
	   This will be  discussed in more
	   detail below but, first, Bell's position on the EPR experiment 
  requires an explanation.
	    
\subsection{Bell's theorem}

Certainly, the most significant development arising out of the prolonged
 debate on the EPR experiment  is
 Bell's theorem  \cite{NewBell,Bell1}.  However, even though its importance is 
never disputed, what 
the theorem actually proves remains a  matter of
 controversy (see, {\em e.g.}, \cite{Norsen1}). The discord can be attributed, in part, to
an ``assumption'' that Bell makes in his theorem; namely, the infamous 
``hidden variables''. These
being the variables that Bell includes in addition to those used in the standard description of quantum mechanics.

We will, for current purposes, turn  to a modern interpretation of Bell's work, as delineated in a series of articles by Norsen
\cite{Norsen3,Norsen1,Norsen5,Norsen4,NewNorsen}.
      This account of the  theorem utilizes Bell's description of locality as the primary  starting point. (Bell's locality definition is clarified
in Section~2.) One of Norsen's main  points of emphasis is
that Bell did not intend for his analysis to serve
     as a  stand-alone argument. Instead, the EPR argument must be considered as the first  of a two-step procedure, with the second step being the
formulation of  Bell's celebrated inequality.
      This is a nuance that has been  missed by some, leading
to a misunderstanding
      about the significance of Bell's ``hidden variables''. 
      All the relevant variables, whether hidden or otherwise, are meant to be
 contained in the initial-state parametrization, $\lambda$.
      Therefore, the inclusion  or exclusion of hidden variables is
really  besides the point when Bell's inequality is implemented.       
   Consequently,   what  Norsen's treatment  shows is how any theory, with or without hidden variables, that  is able to correctly describe the behavior of
      quantum particles must (in some cases) violate  Bell's inequality
and, therefore, disobey his criteria for locality.
      
This  argument will be reviewed, in due course, with particular emphasis on
 the role that is  played by  counter-factual statements; 
these being  ubiquitous in
Bell's formulation.  Such statements are, in this context, referring to hypothetical  measurements that could have been performed by Alice and Bob but are never
actually carried out. 
      Following this review, we will present a different approach, from the perspective of 
observer complementarity,
    that  resolves any issues regarding   ``spooky action at a distance''.
      	  
\subsection{Nature's censorship of paradoxes}

The principle of observer complementarity arose out of
  horizon complementarity,
 which applies
to Nature's seemingly exotic behavior around the horizon of a black hole 
\cite{Suss3,Hooft,Suss2,Bousso}. In either case,  the principle
is that, although different observers can disagree on the occurrence
of certain events,
 Nature will not allow any  observer to ever experience a paradoxical
situation. In order to understand the basic argument,
the notion
of information flowing out of a  black hole must first be understood. 
      
It was Hawking who first showed that black holes slowly evaporate by 
emitting radiation with a nearly thermal (black-body) spectrum  \cite{Hawking}.
Importantly,  this effect
  can be attributed  to the influence of the black hole's 
gravitational field  on {\em quantum}  matter fields; that is,  black holes do 
not radiate unless quantum effects are accounted for.
      Therefore, black hole radiation is realized in a situation for which  
{\em both} gravity  and quantum theory are important.
      This process, however, posed problems for
the  quantum side  as a pure state entering
      the black hole will later be observed  in the radiation as a 
mixed state; information is apparently lost \cite{Hawking2}.
      Hawking later revised this position \cite{Hawk}, as the viability of information loss was challenged by 
      ideas from string theory, the tentative theory of quantum gravity 
\cite{Mald}.
      These ideas suggest that any process which results in information loss 
is strictly forbidden in what is 
a  manifestly unitary theory (namely, the quantum-field-theory dual to 
Einstein's gravity \cite{Mald2}).

      Hawking's original calculation used only an approximation of  quantum gravity 
 because it
      described the black hole as a classical body. In recent work,
      a more rigorous version of Hawking's calculation, using a description of 
the black hole
      in quantum terms, reveals that  information is {\em not} lost
\cite{Med} ---
 much in the 
      same way that information from a burning encyclopedia can, in principle, be recovered.
These arguments are beyond the scope of the present
      paper, but what is essential is that  information
can indeed be accessed from the radiation.
      The availability of this information to an observer outside of the black hole is the starting point
      of Susskind's standard argument for horizon complementarity 
--- see \cite{Susskind} for a simplified account.

	    To illustrate this argument, two observers are considered by Susskind; one falling freely into the black hole
	     and one 
 watching the black hole from afar. 
	    When the first observer, Alice, falls past the horizon of
the black hole,~\footnote{The horizon represents the   
``surface of no return''; classical matter can never escape from the black hole
interior after  passing through this causal boundary.}  
she is 
	    seen by the distant observer, Bob, as being vaporized
before ever reaching the horizon. This is because, from Bob's perspective,
the near-horizon region is extremely hot.
	  Then,  from the above description, it can be deduced that
	    the ``Hawking radiation'' will eventually contain information about 
Alice that can be accessed by Bob. 
	    	    
	    However, Alice, from her point of view, experiences no change in scenery as the space near  the horizon of a large enough black hole
	    is almost flat.~\footnote{The two disparaging  descriptions 
are fully consistent with general relativity and
can be attributed to the  large gravitational red-shift 
between the two different perspectives.}~\footnote{We are overlooking
the recently posed possibility of a sea of high-energy quanta,
 a  ``firewall'',  obstructing Alice's
passage through the horizon \cite{AMPS}, as the validity of
this claim is still actively being contested; {\em e.g.}, \cite{Pap}.}
 And so we are left with a situation in which one or the other
	    experiences a violation of the laws of nature; either Alice 
thermalizes instead of 
	    experiencing flat space or Bob observes Alice falling in
without thermalizing.
	    Susskind's next step is to consider whether either observer {\em must \/} see a violation. An apparent  resolution is to assume that
	   the  information (or Alice)   is ``cloned'' at the horizon,
so that one
	    copy of Alice is radiated out and the second copy  falls in. However, this requires violating a basic
	    principle of quantum mechanics: Quantum information can not be duplicated
in this way  since it would be in conflict
	    with the principle of  linear superposition  \cite{zurek}. 
	    
	    Considering the two observers, Susskind clarifies that the
key to resolving this problem  lies with the movement of information. Alice is safe from seeing any
	    violation since she is trapped within the black hole horizon but Bob has the opportunity
	    to gather Hawking radiation (and hence information) from the black hole and
	    then follow Alice in. If, after entering the black hole, Bob could receive a
	    signal from Alice, there would be a violation of Nature due to 
Bob having  observed two   cloned
	    copies of the same  information. 
Susskind, however,  presents    an argument which proves that
this is not a problem.  

After passing through the horizon,  Alice has a limited time to send 
a signal to Bob {\em before} she  hits the singularity
of the black hole
(where all matter would be  destroyed by immensely large tidal effects). 
What Susskind shows --- by applying the Heisenberg uncertainty principle 
and the knowledge that Bob has to wait a certain time
before he can retrieve a  copy of the  signal from the Hawking radiation
 \cite{Page,HP} ---   is that 
Alice's signaling device would  then  have
	  to be   more energetic than the  black hole itself.
Consequently, Alice's device would no longer be able to fit inside the 
black hole 
\cite{bek}, rendering the whole experiment as moot.
	    And so we see that each observer has an individual account of events, which
	    differ, but that neither observer can possibly compare these results and create a paradox;
	    Nature simply does not allow it.  

      This idea is very relevant to the EPR problem. Horizon complementarity has since been
      molded into observer complementarity, which promotes the 
former principle to one having more general applicability   \cite{Bousso}. Since horizon complementarity
      arises as a consequence of the synthesis of   gravity and
quantum theory,  it appears to be a principle
of quantum gravity. 
As the (presumed) fundamental theory of physics, quantum gravity  
      should be considered to be where all other theories emerge from, and so
      certain aspects of quantum gravity theory will apply to these emergent 
theories \cite{Butter}. Certainly, not all
      aspects of quantum gravity will apply to all emergent theories,  but there is no good reason not to consider the
      application of observer complementarity to, say,
 standard quantum mechanics. 
If an otherwise paradoxical situation can
      be resolved by observer complementarity, its usage can then be justified
{\em a posteriori}.

      The above argument demonstrates that each observer of the black hole will have 
his or her own distinct account of the events having transpired. 
The expansion of this idea
      is that a theory describing the experiences of two or more observers must account for only one observer's results
      at a time. A collective description of two (or more) experiments that are not causally connected can
      lead to paradoxical results and, if so,   each experiment must then be 
described individually.
   We will apply this very idea  to the EPR problem, allowing each observer
to have  his or her  own description of the situation.

\subsection{Similar stances}

There are some approaches in the literature  which are similar to that of the current paper
but  with different motivations.

    One approach is that applied by Mermin in his so-called 
Ithaca interpretation of quantum mechanics \cite{Mermin1,Mermin2}.
This viewpoint places its conceptual  emphasis on  the correlations between the constituent
subsystems of the total quantum system. What Mermin shows
is that these correlations are entirely  captured by the system's
density matrix and can be revealed
by suitable tracing procedures.  He then argues that this is the 
correct framework
for describing reality in the quantum world.~\footnote{Mermin
originally asserted that  correlations between subsystems provided
a complete description of quantum reality but has since retracted
this claim \cite{NewMermin}.} Our stance is similar because, as seen
later, applying observer complementarity is tantamount
to tracing over the inaccessible variables of the density
matrix.

    Another such approach is  that of  ``relational'' quantum mechanics,
 as first  presented  by Rovelli \cite{Rovelli2}.
 This interpretation  is 
    founded on the idea of describing reality
strictly  in terms of relations between (quantum) observers.
This is philosophically similar to but operationally distinct from observer
complementarity. Indeed, Rovelli and Smerlak's resolution
of the EPR paradox \cite{Rovelli2} resembles the current presentation;
nonetheless,
our motivation will be focused on adhering to the requirements of
    observer complementarity without resorting to additional assumptions and inputs
from outside the realm of standard quantum mechanics.

  Another common link between our treatment and Rovelli's
is with regard to  the concept of a ``super-observer''.
    By assigning an  element of reality  to Alice's
    prediction of what Bob measures (or {\em vice versa}), EPR requires a hypothetical observer that can  ``see'' the outcome of the prediction
even if the implicated measurement never actually happens.
    Essentially, the predicted value must exist for some hypothetical observer who has access to all
    information that is held in the Universe.
    This element of the argument is elaborated on  later in Section~4.

\section{Bell ({\em ala\/} Norsen)}

In order to better appreciate our observer-complementarity approach to the 
EPR paradox,
we will first outline the results of Bell's mathematical treatment.
The crux of Bell's argument lies in the formulation of his locality
condition.  Understanding this definition correctly  
is crucial to realizing the significance of  Bell's inequality, and a diversity 
of locality definitions amongst authors has led to a divergence of opinions
regarding the theorem and its results. Norsen, in particular, presents
Bell's theorem by using a definition of locality
that he  refers to  as ``Bell locality''
\cite{Norsen3}.

The condition relies mostly on Bell's original statements regarding
a pair of  space-like separated observables. 
The requirement of locality 
is that the probabilities associated with one of the observables, when worked 
out from
a complete description of this one's past interactions, does not rely
on the other, space-like separated observable. In other words, the information at Bob's
location must be irrelevant to the probabilities being calculated at Alice's location (and {\em vice versa}). 

And so, in terms of the EPR experiment, this locality condition translates
into 
\begin{align}\label{locality}
P(A|\hat{n}_{a},\hat{n}_{b},B,\lambda)\;&=\;P(A|\hat{n}_{a},\lambda)\;,
\end{align}
where $P$ is the probability of Alice's outcome being $A$, conditioned
 by the specified variables on the right-hand side of the vertical
divider.
Also, $B$ refers to the value measured by Bob, $\hat{n}_{a}$
and $\hat{n}_{b}$ refer to Alice and Bob's respective choices of measurement direction, and 
$\lambda$ represents a complete description of the original singlet state
as  prescribed by the theory
under scrutiny \cite{Norsen1}. 
 
Equation (\ref{locality}), the mathematical statement of Bell's
criteria for  locality, does not permit 
any non-causal action to influence the separated observations.
It says that  the probability of result $A$ 
must be the same whether values from Bob's location are taken into account or not, as any dependence
of $A$ on Bob's outcome must be excised due to their acausal (space-like) 
separation. Put differently, Bob's outcomes
are non-local with respect to
 Alice's. 

Adjoint to this locality definition is the requirement of separability:
The joint probability for Alice and Bob must factorize
into the product of two separate probabilities, one for each observer 
individually \cite{Norsen3}.
This factorization should, as Bell argued \cite{Bell1}, be considered as a consequence of
the  locality condition rather
than a new input. 
The conceptual motivation for this being that the
 locality condition~(\ref{locality}), as well as its $A\leftrightarrow B$ converse, requires each observer to have
independently calculated probabilities at their respective locations. 
But the joint probability for the singlet state 
must still  be preserved,  and so this expression 
should be  separable  into a pair of  independent probabilities.
The mathematical description of this factorization, in terms of the Alice--Bob setup, is then
\begin{align}
P(A,B|\hat{n}_{a},\hat{n}_{b},\lambda)\;&=\;P(A|\hat{n}_{a},\lambda)\cdot{}P(B|\hat{n}_{b},\lambda)\;.
\label{factor}
\end{align}
Notice that, in addition to   separating the joint probability
on the left, we have applied 
Bell locality  to each observer's probability statement.
Hence,  each respective probability depends only on
the  locally chosen direction of measurement, $\hat{n}_{a}$ or $\hat{n}_b$,
and information pertaining to the initial state of the  measured particle
 as encoded in  $\lambda$.
It is 
this relation that leads one to the inequality which Bell devised to test theories for adherence to locality.

The real problems emerge when this formalism is
confronted  with  the results from
 actual  quantum experiments. Rather than reiterate Bell's derivation,
we consider a simple example. As known from both experiment and standard quantum mechanics,
the probability of measuring spin up or spin down for a member of a 
singlet pair, in any given direction, is $50\%$. 
Mathematically, this means  for Alice that
\begin{align}\label{alice}
P(A=\uparrow|\hat{n}_{a},\lambda)\;&=\;\frac{1}{2}\;,
\end{align}
where $\lambda$ should now be regarded as
the quantum-mechanical  wave function 
 describing the singlet state.

On the other hand, experimental results also tell us that Alice's measurement
must adhere to the anti-correlation of the singlet, whereby
\begin{align}\label{alicebob}
P(A=\uparrow|\hat{n}_{a},\lambda,\hat{n}_{b}=\hat{n}_{a},B=\downarrow)\;&=\;1
\;, \\
P(A=\uparrow|\hat{n}_{a},\lambda,\hat{n}_{b}=\hat{n}_{a},B=\uparrow)\;&=\;0
\;.
\end{align}
That is, when Bob's chosen measurement direction corresponds with Alice's 
choice, $\;\hat{n}_{b}=\hat{n}_{a}\;$, the spin that Alice measures
must be the opposite of Bob's.

When these two outcomes are compared with the condition
of Bell locality, 
one can see a clear violation. The locality condition demands that 
$\;P(A|\hat{n}_{a},\hat{n}_{b},B,\lambda) = P(A|\hat{n}_{a}, \lambda)\;$.
When $\;\hat{n}_{a}\ne\hat{n}_{b}\;$, there is no such violation; for instance, the  substitution of  (\ref{alice}) and (\ref{alicebob})
into the locality condition~(\ref{locality}) gives
\begin{align}
 P(A=\uparrow|\hat{n}_{a},\lambda)=\frac{1}{2} &= \frac{1}{2}=P(A=\uparrow|\hat{n}_{a},\lambda,\hat{n}_{b}\ne\hat{n}_{a},B=\downarrow)\;. \label{inequal} 
\end{align}
However,  when
$\hat{n}_{a}=\hat{n}_{b}$, the  same substitutions  
rather yield
\begin{align}\label{inequality}
 P(A=\uparrow|\hat{n}_{a},\lambda)=\frac{1}{2} &\ne 1=P(A=\uparrow|\hat{n}_{a},\lambda,\hat{n}_{b}=\hat{n}_{a},B=\downarrow)\;.
\end{align}
This inequality shows that, in order
to respect Bell locality, quantum mechanics must include more than what is
currently held in the wave function.
It should be  noted that  the  above outcome represents only one particular case of Bell's 
more general mathematical statement, his celebrated inequality \cite{NewBell}.

The inequality within~(\ref{inequality}) would  seem to imply that quantum mechanics
is simply incomplete as a theory and, in order to fully describe reality,
 extra  or  ``hidden'' 
variables are required.
However, Norsen's treatment 
shows that {\em any \/} theory which adheres to Bell locality, with or without
these hidden variables,~\footnote{Bohmian quantum mechanics is
 one such example of a hidden-variable theory  \cite{Bohm}.} cannot explain the perfect anti-correlation which
is verified by experiment \cite{Norsen1}. This is because any such 
variables can be included {\em a priori\/}  in $\lambda$ and, thus,
 lead to the very same conclusions.
And so, in view of this argument, one is forced
either to reject any adherence to Bell locality within quantum mechanics
{\em or} to accept quantum mechanics only as a calculational device that
does not fully describe reality.

The crucial part of Bell's argument, as far as this paper is concerned, is
the emergence of {\em counter-factual definiteness} (CFD),
as demonstrated and elucidated  by Norsen \cite{Norsen1}.
CFD is the claim that a statement about a measurement which was {\em not\/} performed
can be discussed, in a meaningful way, alongside statements about actually performed experiments.
This can be seen in (\ref{inequality}), where the left-hand side of the expression
assumes no knowledge of Bob's measurement. This is equivalent to Bob
not yet having performed
any measurement, as the left-hand side is not conditioned
by  any action taken by Bob. However,
the right-hand side of (\ref{inequality}) assumes Bob {\em did\/} perform a measurement
and found a particular outcome which influences Alice's result. 

In order to illustrate this idea, let us consider the situation of placing a bet at a 
roulette wheel. A pessimistic gambler might suggest a ``theory'' that, if a bet is placed on red,
the wheel will stop on black, otherwise it will stop on red.
Perhaps,  there is a  particular spin of the wheel
that  substantiates the gambler's
(flawed) assertion.
However, it cannot be claimed, after the fact, that a change in the bet
would have reversed the outcome on the wheel.
Such a claim constitutes  a discussion
of what could have happened, but {\em did not\/}, in lieu of  what
actually {\em did}.

This requirement of CFD within Bell's theorem is an inherent part of the
overall  argument
and cannot be arbitrarily eliminated --- it  arises as a direct {\em consequence} of Bell locality. 
The fact that assuming Bell locality necessarily implies the use of CFD is
clearly demonstrated by Norsen  \cite{Norsen1}.
Moreover, the same basic claim can be made for the stochastic or probabilistic reformulation of Bell's inequality,
which is known as the Clauser--Horne--Shimony--Holt  inequality \cite{CHSH}.~\footnote{This entails that the state description be averaged over all possibilities: \;$|\lambda\rangle \rightarrow \int{}\rho(\lambda)|\lambda\rangle{}d\lambda\;$,
where $\rho(\lambda)$ is a weight factor such that $\;\int{}\rho(\lambda)d\lambda=1\;$.
So that, for example, $\;P(A=\uparrow|\hat{n}_{a}, \lambda) \rightarrow \int{}\rho(\lambda)P(A=\uparrow|\hat{n}_{a}, \lambda)d\lambda\;$.}
While probabilistic theories do remove CFD, they retain a slightly weaker form,
counter-factual meaningfulness (CFM), which results in the same dependence on
discussing events which could have happened but did not \cite{Norsen5}. 
These CFD and CFM statements are, not only inevitable in the construction of
Bell's inequality, but will be a central element in our  
proposed resolution for alleviating the  concerns of EPR and Bell.

\section{Reworking EPR and Bell}

We begin here  with the same setup as previously considered: 
There is a prepared singlet state, as in (\ref{singlet}),
consisting  of two particles, $\alpha$ and $\beta$.  Each of the particles is 
sent to one of a  pair of spatially separated observers, Alice and Bob, who
 are then free to measure the
spin of their particles in the direction  of their choosing. Alice will be
measuring the spin of $\alpha$, which is
given by $\;\vec{S}_{\alpha}\cdot\hat{n}_{a}=\pm\hbar/2\;$,  and
likewise for Bob,
$\;\vec{S}_{\beta}\cdot\hat{n}_{b}=\pm\hbar/2\;$.

 Due to the properties
of the singlet state, if Alice and Bob choose to measure the spin
in different directions, then their results will have no correlation.
For the sake of this argument, we are concerned  with the case
in  which Alice and Bob measure along the same spin direction, as this is
where the ``problems'' arise. We are assuming that they never plan to measure
the same direction but  end up doing so in the trial of immediate interest.
 Hence, this is {\em not} a description
of a conspiracy between the observers to measure in the same direction.

Let us recall the singlet state~(\ref{singlet}),
\begin{align}\label{singlet2}
 |\Psi_{singlet}\rangle &= \frac{1}{\sqrt{2}}\Big[|a+\rangle|b-\rangle-|a-\rangle|b+\rangle\Big]\;.
\end{align}
Given the coincidence of measurement directions,
$\;\hat{n}_a=\hat{n}_b\;$, the notation
can now be reinterpreted as meaning
 $\;a\pm=\vec{S}_{\alpha}\cdot\hat{n}_{a}=\pm\hbar/2\;$ and 
$\;b\mp=\vec{S}_{\beta}\cdot\hat{n}_{b}=\mp\hbar/2\;$.  
This is only a change of basis; the state has not been altered.
 
 So far,  we have been  describing  the particle system. 
To talk about {\em observer} complementarity, what  we really need to 
look at
is  the particle--observer system. 
In this regard, a subtle point is that Alice and Bob cannot directly measure the spin
of their respective particle. 
Each of their  measurements involves some (other)
observable, the measurement device,  with its corresponding  value  indicating the result of the
experiment. Let us denote this observable by $|X_{ready}\rangle$ and 
conceptualize it as a pointer that begins in the horizontal position, facing zero.
Then $|X\uparrow\rangle$ and $|X\downarrow\rangle$ will denote a
measurement of spin up or spin down, respectively.

We now let $|A\rangle$ and
$|B\rangle$ respectively represent the result of Alice and Bob's
measurements.  The system for the  particles {\em and} observers
($P$-$O$) can then be described  as
\begin{align}
|\Psi_{P-O}\rangle&=\frac{1}{\surd2}\Big[|a+\rangle|A_{ready}\rangle|b-\rangle|B_{ready}\rangle-|a-\rangle|A_{ready}\rangle|b+\rangle|B_{ready}\rangle\Big]\;,
\end{align}
where the measurements have not yet been performed.

From here, we can calculate
the density matrix of the combined system,  $\;\rho_{P-O}=|\Psi_{P-O}\rangle\langle\Psi_{P-O}|\;$, giving
\begin{align}
 \rho_{P-O} = \frac{1}{2}\Big[&|a+\rangle|A_{ready}\rangle|b-\rangle|B_{ready}\rangle\langle{}a+|\langle{}A_{ready}|\langle{}b-|\langle{}B_{ready}| \\ \nonumber
   -&|a-\rangle|A_{ready}\rangle|b+\rangle|B_{ready}\rangle\langle{}a+|\langle{}A_{ready}|\langle{}b-|\langle{}B_{ready}|  \\ \nonumber
   -&|a+\rangle|A_{ready}\rangle|b-\rangle|B_{ready}\rangle\langle{}a-|\langle{}A_{ready}|\langle{}b+|\langle{}B_{ready}|  \\ \nonumber
   +&|a-\rangle|A_{ready}\rangle|b+\rangle|B_{ready}\rangle\langle{}a-|\langle{}A_{ready}|\langle{}b+|\langle{}B_{ready}|\Big]\;.
\end{align}
The density matrix provides us with a description of all the possible outcomes once the measurements
occur. It should be stressed, though, that this density matrix is not attributed to any
of our observers. It contains information regarding two spatially separated
locations; namely Alice and Bob's measuring stations.
 In order to adhere to causality, we consider only local
sources of information. For instance,  since Alice is separated
from Bob, her description must not access any   information
that is localized at  Bob's station.  So that,  to find Alice's measurement outcomes,
we must first determine the {\em reduced} density matrix relating to her experiment.

Given a quantum density matrix $\rho$, it is standard operating procedure to
remove what a given observer does not know by tracing out the 
hidden systems \cite{Dirac}.
Mathematically, this entails calculating  $\rho_{reduced}=\sum\limits_{a'}\langle{}a'|\rho|a'\rangle\;$,
where $|a'\rangle$  collectively represents the  state kets for 
all the concealed systems. For Alice, this translates into 
$\;\rho_{Alice}=Tr_{Bob}[\rho_{P-O}]=\sum\limits_{b}\sum\limits_{B}\langle{}b|\langle{}B|\rho_{P-O}|B\rangle|b\rangle\;$.

Therefore,  Alice's results are described by~\footnote{We are assuming
orthonormal sets of bases, even for the measuring devices.}
\begin{align}
\rho_{Alice}&=\frac{1}{2}\Big[|a+\rangle|A_{ready}\rangle\langle{}a+|\langle{}A_{ready}|\;+\;|a-\rangle|A_{ready}\rangle\langle{}a-|\langle{}A_{ready}|\Big]\;,
\end{align}
and, for Bob, we find that
\begin{align}
 \rho_{Bob}&=\frac{1}{2}\Big[|b-\rangle|B_{ready}\rangle\langle{}b-|\langle{}B_{ready}|\;+\;|b+\rangle|B_{ready}\rangle\langle{}b+|\langle{}B_{ready}|\Big]\;.
\end{align} 
\par

Let us briefly pause to consider the  measurement process.  For the
sake of explanation, let us (following \cite{Adami}) consider a device
which
measures the position of a  particle.
Given an initial  state  $\;|x\rangle\;$ for the particle 
and  $|A\rangle$ for the measurement device, we can construct a 
description of the combined system
as $\;|\Psi\rangle=|x\rangle|A\rangle\;$.  We can also define a unitary operator for the combined system ---  the interaction Hamiltonian,
$\;\hat{H}_{int}=\hat{X}\hat{\Pi}_{A}\;$. Here,
$\hat{X}$ is the position operator for the particle and
$\hat{\Pi}_{A}$ depicts  a ``momentum'' operator
that is associated with  the measuring device.  Formally,
$\hat{\Pi}_{A}$ is the canonical conjugate to $\hat{A}$ 
and, conceptually, it  is  the operator which
enables   $|A\rangle$ to record the measured result. Assuming a completely
efficient measuring device, we can, without describing it explicitly,
denote a measurement as 
\begin{align}
|\Psi\rangle&\rightarrow e^{i\hat{H}_{int}}|\Psi\rangle \\ \nonumber
&=\left[e^{i\hat{X}\hat{\Pi}_{A}}|x\rangle\right] |A\rangle 
\\ \nonumber	    
&=|x\rangle{}e^{ix\hat{\Pi}_{A}}|A\rangle \\ \nonumber
	    &=|x\rangle|A+x\rangle\;, \nonumber
\end{align}
where $\hat{X}$ and $\hat{\Pi}_A$  are  assumed to commute. 
And so  the device has successfully registered the position
of the particle. This is a very simple example. 
The general case for quantum mechanics involves 
$\;|x\rangle\rightarrow|x^{\prime}\rangle\;$.
This is due to the quantum state of the particle
being changed by the interaction; the particle does not necessarily
end up in its original state.

Applying this basic idea of measurement to Alice's reduced density matrix, we obtain
\begin{align}
\rho_{Alice}&=\frac{1}{2}\Big[|a+\rangle|A\uparrow\rangle\langle{}a+|\langle{}A\uparrow|\;+\;|a-\rangle|A\downarrow\rangle\langle{}a-|\langle{}A\downarrow|\Big]\;
\end{align}
and analogously for Bob. Importantly, there is perfect correlation
between the pointer direction and the spin of the particle. 
Notice that the information from Alice's measurement coupled with  knowledge of the initial state would enable her to predict the spin of $\beta$ in this same
direction. But this can only be a prediction because of Alice
being spatially separated from $\beta$.
There would need to be some {\em local} interaction between Alice and $\beta$ (or Bob) to
provide an actual observable that she could measure.

In order to ensure that quantum-mechanical consistency holds, we have to verify that
the correlation of the singlet state is preserved when Alice and Bob's notes are compared.
To do this, we follow the example of \cite{Rovelli1} and consider a third observer who checks 
up on Alice and Bob's results after their measurements. 

First, let us formulate the density matrix for the combined systems {\em after} 
the measurements,
\begin{align}
 \widetilde{\rho}_{P-O}=\frac{1}{2}\Big[&|a+\rangle|A\uparrow\rangle|b-\rangle|B\downarrow\rangle\langle{}a+|\langle{}A\uparrow|\langle{}b-|\langle{}B\downarrow| \\ \nonumber
    -&|a-\rangle|A\downarrow\rangle|b+\rangle|B\uparrow\rangle\langle{}a+|\langle{}A\uparrow|\langle{}b-|\langle{}B\downarrow| \\ \nonumber
    -&|a+\rangle|A\uparrow\rangle|b-\rangle|B\downarrow\rangle\langle{}a-|\langle{}A\downarrow|\langle{}b+|\langle{}B\uparrow| \\ \nonumber
    +&|a-\rangle|A\downarrow\rangle|b+\rangle|B\uparrow\rangle\langle{}a-|\langle{}A\downarrow|\langle{}b+|\langle{}B\uparrow|\Big]\;.
\end{align}
 Again, it should be pointed out that
this density matrix is not attributed to any observer in our system. Rather, it is a  description that could only
be utilized by  a super-observer having access to all the spatially separated situations. This is
in direct conflict with observer complementarity and so it must be made clear that we do not consider
this as a density matrix that can viably be compared with those of the different observers. 

We will rather
use this matrix to determine  the reduced density matrix for the third observer,  Carol, by tracing out what she has no access
to; namely, the initial states of $\alpha$ and $\beta$. This entails
computing 
\begin{align}
\rho_{Carol}&=\sum\limits_{a'}\sum\limits_{b'}\langle{}a'|\langle{}b'|\widetilde{\rho}_{P-O}|b'\rangle|a'\rangle\;,
\end{align}
where $\;a'=\{a\pm\}\;$, $\;b'=\{b\pm\}\;$ and,
by tracing over the
variables separately, we are not assuming any  (anti-) correlations.
 The outcome is
\begin{align}
\rho_{Carol}&=\frac{1}{2}\Big[|A\uparrow\rangle|B\downarrow\rangle\langle{}A\uparrow|\langle{}B\downarrow|\;+\;
 |A\downarrow\rangle|B\uparrow\rangle\langle{}A\downarrow|\langle{}B\uparrow|\Big]\;,
\end{align}
which shows that the anti-correlation is preserved from Carol's perspective. 

And so we see that this
series of local interactions, in compliance with observer complementarity and
causality, holds no paradoxes.
Any ``spooky action at a distance'' disappears when  the view of each
observer is restricted to  his or her own reduced  density matrix.

\section{Discussion}

\subsection{Against counter-factual definiteness}

 The above  results  show that observer complementarity can account for
the view of  each respective observer,  while still maintaining the 
anti-correlations between the particles as evident from Carol's perspective.

 On the other hand,  Norsen's formulation of Bell's theorem
 proves that  CFD is a necessary consequence of assuming
Bell locality \cite{Norsen1}.
 This can be weakened to CFM for a probabilistic theory but it still 
amounts to the same basic result.
 The theorem inevitably depends on discussing events that  have no associated observer
 because, even though they {\em could\/} have happened, they {\em did not\/} actually happen.
 The idea of discussing, in a meaningful way, a statement about a hypothetical event
 is in opposition to the principle of observer complementarity. From
 the vantage point of this principle,
 an observer has no need to account for the results of an experiment that cannot be compared to his or her own findings. In other words, Alice {\em need not\/} account for
 Bob's results (and {\em vice versa}) until they can be locally compared.
Similarly,
 results from an experiment that  was not actually performed can have no meaning
in the observer-complementarity framework.
It follows that  CFD is an assumption that cannot be
 made in conjunction with observer complementarity;
these being  antithetical constructs. To show that one
 is true in the domain of a given theory  would be
enough  to rule out the
 other within the same realm.
 
 To this end, observer complementarity falls back on its origin. As already 
stated, 
 quantum gravity can be considered as the fundamental theory from which all others are emergent.
 Therefore, as observer complementarity survives quantum gravity, it is the CFD assumption
 which ultimately must fall away. Then, in the spirit of  {\em Occam's razor},
 the simplest way
 to relieve the tension between CFD and observer complementarity is to discard CFD in any paradoxical  situation. 
 This would imply that, due to the inevitable  appearance of CFD, Bell's
 theorem is doomed from the start. 
With the assumption of  Bell locality and, in the case of EPR, 
 their criterion for reality, the respective arguments 
 set themselves up to produce results that will imply either that Nature is non-local or
 that quantum mechanics is incomplete. 
 However, considering the same situation without assuming Bell locality and, instead,
 using observer complementarity as the motivating principle,
we see that  the concerns fall away without forcing 
 a rejection of either option. No contradiction is experienced, as shown by Carol's results,
 and all action happens locally.
 This, along with the fact that observer complementarity is a consequence 
 of quantum gravity, suggest that Bell locality and, by implication,  CFD are 
too strong to define reality.

\subsection{Against conspiracy}
 
 One might still be concerned as to how  the particles ``know'' in advance  which directions will be chosen by Alice and Bob. 
It becomes problematic  if this information is
not  provided by the 
 theory because the only other alternative is conspiratorial settings by 
Nature.  
That is,   
 Nature would, somehow, be anticipating the choices of the observers and adjusting the particles 
 accordingly but without allowing access to the determined values.
 
 To address this question, 
let us consider the evolution of the system through time.~\footnote{Similar sentiments to the following discussion  have appeared  in \cite{NewHooft,Stoica}.}
It should be clear that the initial conditions of a system and its subsequent
 evolution produce the final conditions. But is the converse true? Can
we not evolve the system back in time if given the final
 conditions? For a deterministic or classical environment,
 this is obviously acceptable. 
The quantum realm is not so deterministic, and so the answer is less clear. 
But, perhaps surprisingly,
 the proposition of evolving a quantum system  back in time
  is valid as well  \cite{aharonov}. Indeed,  the choice of final-state
boundary conditions
 does not violate causality nor unitarity, and  so we are
 free to apply reversed time evolution to our previously described experiment.

Let us then consider the  backwards evolution from a time $t_{1}$, when Alice has
 measured a $+$ spin on her particle in the direction $\hat{n}_{a}$.
We can evolve 
this state 
 backwards to a time $t_{0}$, just when the 
 particles were separated.
 As  particle $\alpha$ experiences no interactions during the intervening period, its time evolution is trivial,
 \begin{align}
  |\hat{n}_{a},+\rangle{}_{t_{0}} &= e^{i\hat{H}_{free}(t_{0}-t_{1})}|\hat{n}_{a},+\rangle \\ \nonumber
  &= (phase)|\hat{n}_{a}, +\rangle\;,
 \end{align}
where we have used that the free-particle Hamiltonian commutes with
all spin operators.
Since, at time $t_{0}$, the particles are known to be anti-correlated,
 it can be deduced that
\begin{align}
\label{newstate}
 (phase)|\hat{n}_{a}, +\rangle|\hat{n}_{b}=\hat{n}_{a},-\rangle
\end{align}
describes the state of the two particles.

Hence, as far as Alice is concerned, there has been no conspiracy, and the
same is true for Bob. But what about the initial holder
of the singlet state? After all, the state~(\ref{newstate}) differs
from the original  singlet state~(\ref{singlet2}), and we cannot
appeal to observer complementarity in this case since the initial
location is causally linked to  both Alice nor Bob.
Nevertheless, what can be appealed to is that the wave function depends
on an observer's choice of ``gauge'', much in  the same way that the electromagnetic potentials are gauge-dependent  fields.  That is, even causally connected
observers need only agree on gauge-invariant, physically measurable quantities.
Here, all observers agree that the spins are anti-correlated, which
is all that can ever be known with certainty.

There may also be concerns that the quantum particles
do not ``perceive''  the arrow of time and, consequently, have the  capability
of time travel into the past.
It is, in fact,  already known that 
final-state selection does allow for such time travel  via
the process of  post-state
teleportation; an idea that is discussed at length  \cite{tt1}.
While this is an area of study with many open questions, two important
points  will be made to assuage any discomfort at the suggestion of
this type of particle behavior. (See \cite{tt1} for further details
and references.)

Firstly, this form of time travel  can be formulated
in away that does not jeopardize the standard tenets
of quantum mechanics, nor does it
lead to any  stereotypical ``grandfather
paradoxes''.~\footnote{This refers to the prototypical paradox of time travel: 
That a person could go back in time and
 shoot his or her own grandfather before  ever being conceived.}
And so this concept can safely be applied to quantum particles without 
worrying about any disorder permeating to the classical realm.

Secondly, given a consistent theory of quantum gravity, it is natural
if not necessary that time travel be incorporated into the quantum side
of the theory at some level. This can be understood as follows:
In the classical
(Einstein) theory of gravity, closed time-loops or wormholes cannot be ruled out by any formal proof.
For instance, the conditions around black holes produce effects which suggest that
there are such time-ambiguous regions.
Therefore, a consistent theory of quantum gravity requires that the quantum
realm   also allows for some notion of going backward in time. Post-state
teleportation can be viewed as a tangible realization of
this requirement.

\subsection{Against super-observers}

As alluded to earlier,
the underlying premise of EPR's criterion for reality, as well as the application of
Bell locality, utilizes an assumption regarding a ``special'' observer with
the  capacity
to view the Universe in its totality; the super-observer. This is a concept 
that has been used throughout
science to  build theories for explaining the world \cite{7}. 

A useful approach to understanding this idea is through ``Laplace's demon''. This is a 
conceptual device that is  used to represent the Universe as a 
deterministically working model  \cite{4}.
It entails a hypothetical super-intelligence having access to
all laws governing the Universe as well as a precise description of the Universe
for a particular moment of time \cite{10}. So that  Laplace's demon, this hyper-intelligent super-observer,
would have accurate knowledge of all values for all things in the Universe.
This, combined with its  knowledge of the physical laws, would allow the prediction of
all future occurrences as well as a retroactive prediction of all  past
events \cite{4}; that is, the ability
to describe the Universe, in its entirety, for all time given only one set of initial values.
Such an entity is then representative of   a 
deterministically functioning universe, 
where it is  only our own  ignorance of accurate values
that  prevents us from perfectly forecasting the future.

And so, for every moment of time, there would be a single
and complete  description of the Universe and all of its constituents. 
This is obviously problematic for observer
complementarity, which does not allow a single observer to have access to all information since he or she cannot have access to any spatially separated values.

The necessity for a hypothetical super-observer has been debated over history.
It was Heisenberg's uncertainty principle which first provided a strong opposition. Insofar as some information is always inaccessible, there is
a fundamental limitation to  the predictive
capabilities of Laplace's hypothetical demon. 
In this way, the workings of quantum mechanics places
a firm restriction on what can be known as a matter of principle. Even if
given all {\em  available} 
information 
at a particular time, the demon would still not be able to consistently make
predictions about quantum events with certainty.

The final nail in the coffin for the super-observer came from a calculation 
that described Laplace's
demon as a computational device \cite{8}. This paper showed that, for a device to accurately predict
everything in a system, it cannot be part of the same system. Essentially, Laplace's demon cannot
be situated within the Universe if it is to provide a total description of the Universal state. 
And so  the conclusion is that, for any set of natural laws, a computational
 device
would not be able to predict everything within a world 
 that it is also contained in. The concept of a Laplacian demon  could still hold  for one  living  ``outside'' the Universe. But such a demon has no 
utility because, in this case,  its predictive power is
then  literally out of reach \cite{9}.
Hence, any theory  could only describe ``almost everything''; there will always be certain
values that are forever out of its reach. (For a recent discussion, see \cite{lloyd}.)

There is, therefore,  no compelling reason to suggest that the contradiction between 
observer complementarity and the  super-observer implies that 
former should be discarded. Rather, the evidence points to the dismissal of 
the super-observer as a {\em part of \/} our Universe. 

\section{Conclusions}

In closing, the approach of  this paper reveals an oversight regarding the assumptions that are used in
the EPR thought experiment  and Bell's formulation thereof. 
As made clear by Norsen and recapitulated above, the  arguments rest 
squarely on the validity of CFD. 
However, as we have pointed out, the application of
 CFD is at odds with the principle of  oberver
complementarity.  
   Then,  since the latter attains its pedigree
from the fundamental 
theory of quantum gravity, we have argued that
Bell locality   is too strong a definition for reality and that this is what leads to the troubling outcomes.
By dismissing Bell locality and utilizing observer complementarity,
 we have shown that
all concerns regarding locality and the completeness of quantum mechanics 
fall by the wayside, as the EPR
scenario naturally resolves itself in a self-consistent manner.

The acceptance of   observer complementarity  also requires one to dismiss
 the notion that a hypothetical super-observer is
present in the Universe. This does not present a problem because, not only is there no argument for
the necessity of a super-observer, there is considerable evidence against 
it.
A super-observer may still fulfill the criteria of Laplace's demon from outside the Universe.  However, 
any interaction of the super-observer with the Universe requires 
it to be part of  the same; in which case, 
Laplace's criteria can no longer be fulfilled. Therefore, any theory explaining 
measurement within the Universe cannot provide a deterministically complete description.

As locality and causality have been maintained by our reworked calculation, 
what must
change is how reality is viewed; not as a single description of all subsystems but, rather, 
as a collection of  descriptions from many different  observer's points of 
view. We would, if it were possible, be inclined to remind Einstein
that such a notion is not much different than his theory of relativity.

\addcontentsline{toc}{section}{Acknowledgements}

\section*{Acknowledgments} 
 
The research of AJMM is supported by NRF Incentive Funding Grant 85353.


\end{document}